\documentstyle[epsfig,nato]{crckapb}

\begin{opening}

\title{THEORETICAL ESTIMATE(S) OF THE CP-VIOLATING \protect\\
       QUANTITY $\varepsilon'/\varepsilon$ IN $K \rightarrow 2 \pi$ DECAYS}

\author{J.O. EEG}

\institute{Department of Physics\\
           P.O.Box 1048, Univ. of Oslo\\
           N-0316 Oslo, Norway}

\end{opening}

\runningtitle{CP-VIOLATION IN $K \rightarrow 2 \pi$}

\def\eq#1{{eq.~(\ref{#1})}}

\def\abs#1{\left| #1\right|}
\def\mod#1{\abs{#1}}

\def\Re{\mathop{\mbox{Re} }}
\def\Tr{\mathop{\mbox{Tr }}\,}

\def\e{$\varepsilon'$ }
\def\ee{$\varepsilon'/\varepsilon$ }
\def\CP{$CP$ }
\newcommand{\bea}{\begin{eqnarray}}
\newcommand{\beq}{\begin{equation}}
\newcommand{\eea}{\end{eqnarray}}
\newcommand{\eeq}{\end{equation}}
\newcommand{\nnu}{\nonumber}

\begin{document}\setlength{\unitlength}{1mm}


\begin{abstract}

I give a short presentation of the 
 theoretical
prediction of $\varepsilon'/\varepsilon$. 
Short distance and especially long distance aspects of the computation 
is discussed. 
I consentrate on the general framework and the chiral quark model
approach, while other approaches 
  are also shortly presented.
Because of the intrinsic uncertainties of 
the long-distance computations,
it is unlikely that new physics effects
can be disentangled from the standard model prediction.

\end{abstract}

\section{Introduction}
The physical neutral kaons $K_{S}$ and  $K_L$ decay predominantly into
two and three pions, respectively. 
The final states  of two and three pions are
even and odd under a \CP transformation, respectively
Therefore, it was natural to think of $K_{S,L}$ as eigenstates 
of \CP -symmetry.
But in 1964 
the \CP -violating decay mode $K_L \rightarrow 2 \pi$ was discovered.
This shows that
 $K_{S,L}$ are linear combinations of the
\CP-parity eigenstates $\sim (K^0 \pm  \bar K^0)$ :
\bea
K_{S,L} \; \simeq  \; \left[ (1+\varepsilon) K^0 \pm 
 (1 - \varepsilon) \bar K^0 \right]/ \sqrt{2} \; \; ; 
\label{KS-KL}
\eea
where $|\varepsilon| = (2.266 \pm 0.017) \times 10^{-3}$.  

While $\varepsilon$ is determined by \CP -violation in the
$\Delta S = 2$ transition 
\newline
 $K^0 \leftrightarrow \overline{K^0}$, 
\CP -violation can also proceed {\em directly} in
 the decays of $K^0$ or $\overline{K^0}$  in two pions 
(the $\varepsilon'$ effect), expressed 
by the  parameters: 
\beq
\eta_{ab} \equiv 
\frac{\langle \pi^a \pi^b | {\cal L}_W | K_L \rangle}
{\langle \pi^a \pi^b | {\cal L}_W | K_S \rangle} \; \; ; \qquad
\eta_{00} \simeq \varepsilon - 2 \varepsilon' \; , \quad
\eta_{+-} \simeq \varepsilon + \varepsilon' \; ,
\label{etaxx}
\eeq
where  ${\cal L}_W$ is the $\Delta S = 1$ weak lagrangian.

In the Standard Model (SM), $\varepsilon$ and \e 
are proportional to the same complex quantity in the
Cabibbo-Kobayashi-Maskawa (CKM) quark mixing  
matrix.  In models beyond the SM there 
might be different  sources of \CP -violation.
 It is therefore of great
importance to establish the experimental value of \e and discuss
its theoretical value within the SM and beyond.

In this presentation, I will consentrate on qualitative aspects.
For technical details I refer to
 reviews \cite{WiWo93,deRaf,Buch96,BEFrev} and recent 
papers \cite{Bosch99,BEFup}
which  also contains the most important references on the subject.

\section{Effective Theory for $\Delta S =1$ Non-leptonic Decays}
Theoretically, the amplitudes for
$K \rightarrow 2 \pi$ depend on different energy scales from
the top quark mass down to the masses of 
the  light $u,d$- quarks. 
In such cases, one  often uses 
{\em effective theories} \cite{Pich}. In this case one
 constructs an effective
quark level  Lagrangian
 \beq
 {\cal L}_{W}=  \sum_i  C_i(\mu) \; Q_i (\mu) \; ,
 \label{Lquark}
\eeq
where all information of the short distance (SD) loop 
effects above a renormalization scale $\mu$
is contained in some (Wilson) coefficients $C_i$.
These depend on the masses of the $W,Z$-bosons,
  the heavy quark masses ($m_q > \mu$) and moreover on
 $\Lambda_{QCD}$ and the 
renormalization scheme.
The $Q_i$'s are quark operators,
typically containing products of two quark currents.
A priori, a huge number of operators are involved, 
 depending on the chosen $\mu$. 

To obtain
a physical amplitude, one has to calculate the hadronic matrix elements
of the quark operators $Q_i$ within some non-perturbative framework.
This is the long distance (LD) part of the calculation.
 Unfortunately,
 lattice gauge theory results are difficult to obtain for some operators.
 Therefore models and assumptions are often used.
 Using quark models, $\mu$ is usually taken below
the charm quark mass. Then the $Q_i$'s contain only
 the lightest three quarks $u,d,s$, and the
dominating operators are considered to be ten operators of dimension 6
(see, however, \cite{CDG}).

For the low-energy sector containing the three light quarks only, there is
a well defined effective theory, chiral perturbation theory ($\chi PT$),
having the basic symmetries of QCD.
One  can try to match these theories by  {\em bosonizing} the quark 
operators $Q_i$:
\beq
 Q_i \rightarrow  \sum_j  F_{ij} \; \hat{{\cal L}}_j \; \; ,
 \label{BosQ}
\eeq
where $F_{ij}$ are quantities which have to be calculated with 
non-perturbative methods (including quark models).
 The $\hat{{\cal L}}_j$'s
are chiral lagrangian terms, (involving $\pi, K , \eta$), to be discussed 
in the subsections 2-4.
Knowing the bosonization in (\ref{BosQ}),
 we could calculate the various $K \rightarrow 2\pi , 3 \pi$ amplitudes
from a $\Delta S = 1$ chiral lagrangian
\beq
 {\cal L}_{W}(\chi PT) =  \sum_j   G_j \; \hat{{\cal L}}_j \;  ; \qquad
 G_j =  \sum_j C_i \;  F_{ij} \; .
 \label{LMes}
\eeq
The challenge within such an approach is that the coefficients should be 
calculated (and matched) in a region where both the SD and LD
calculations  are valid.  
Notice that in pure chiral perturbation theory
 the $\hat{{\cal L}}_j$ 's are known from
symmetry requirements, while the coefficients $G_j$ has to be determined from experiment.

A starting point in many calculations has been the {\em factorization}
hypothesis in cases where  the quark operators contain a product of 
two currents, say. This hypothesis (also named
 ``vacuum saturation approximation'' (VSA))
is easy to handle because the matrix elements of currents are known. 
But corrections to VSA are in general large.

\subsection{The Quark Effective Lagrangian and the  Wilson Coefficients}
The $\Delta S=1$
quark effective lagrangian at a scale $\mu < m_c$ can be 
written
as in (\ref{Lquark})
with
\bea
C_i(\mu) =  - \frac{G_F}{\sqrt{2}}  
 \Bigl[ \lambda_u \, z_i(\mu) + \lambda_t \,  y_i(\mu) \Bigr] \; .
\label{Lqcoef}
\eea
Here $G_F$ is the Fermi coupling, the functions $z_i(\mu)$ and
 $y_i(\mu)$ are the \CP -conserving and -violating parts of the
 coefficients, and  
$\lambda_q = V_{qd}\,V^*_{qs}$ (for $q=u,t$) are the CKM factors.
The numerical values of $z_i$ and $y_i$ 
are of order one down to $10^{-4}$, and can be found in the litterature.
The standard
basis (for $\mu < m_c$) includes  ten  operators. We display four,
which are important for the $\Delta I = 1/2$ rule ($Q_{1,2,6}$)
and \ee ($Q_{6,8}$):
\begin{eqnarray}
Q_{1}  =  4 \,  ( \overline{s}_L \gamma^\mu  d_L )  \; \,
           ( \overline{u}_L \gamma_\mu  u_L )
\; \;  , & \,
Q_{6}  = -8  \sum_{q} \, ( \overline{s}_L q_R ) \; \, ( \overline{q}_R  d_L )
\, , \nnu  \\
Q_{2}  =  4 \, ( \overline{s}_L \gamma^\mu u_L ) \; \,
           ( \overline{u}_L \gamma_\mu d_L )
\; \; , & \,
Q_{8}  =  -12 \sum_{q} \hat{e}_q \; ( \overline{s}_L q_R )                                       \; \, ( \overline{q}_R  d_L )
\, , 
\label{Q1-10} 
\end{eqnarray}  
where  $\hat{e}_q$  are the quark charges 
($\hat{e}_u = 2/3$,  $\hat{e}_d=\hat{e}_s =-1/3$), and $q_{L,R}$
are the left- and right-handed projections of the quark fields. 
Within this basis the Wilson coefficients are calculated 
to the  order  $\alpha_s^2$ and 
$\alpha_s \, \alpha_{em}$ \cite{Bur93,Ciu94}, and it is now 
a basic element
used by all groups estimating \ee.
\begin{figure}
\begin{center}
\epsfig{figure=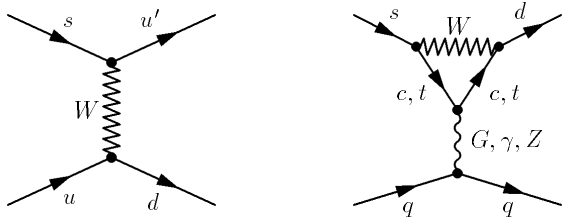,width=10cm}
\caption{$W$-exchange corresponding to $Q_2$, and Penguin diagram(s)}
\label{TrePeng}
\end{center}
\end{figure}

Within the SM, $Q_2$ is induced by $W$-exchange  at the tree level 
(see Fig. \ref{TrePeng}, left).
Switching on QCD, $Q_1$ is induced by one loop corrections to $Q_2$.
The rest of the operators are induced by the penguin diagrams (see
Fig. \ref{TrePeng}, right) and their higher order loop corrections,
$Q_{3-6}$ by gluon ($G$) exchange and 
 $Q_{7-10}$ by photon ($\gamma$) or $Z$ exchange (also box diagrams 
with exchange of two $W$'s enter).

Other possible operators of dimension 5 or 6 exist,
 but give small contributions within the SM~\cite{BEF95,BEsp}.
However, it has been  stated that higher dimensional operators will become
relevant\cite{dim8}  if the renormalization (separation) scale
 $\mu$ is chosen to low\cite{CDG}.

The $\Delta S=2$ case (involving one relevant quark operator only) is treated
 along similar lines.
Note that an analysis of
the  $\Delta S=2$  $K$- $\bar{K}$ mixing determines the CP-violating 
parameter Im $\lambda_t$ in the SM.

\subsection{Chiral Perturbation Theory ($\chi PT$)}
The chiral lagrangian and chiral perturbation theory
provide  a faithful representation of 
the light ($u,d,s$ quark) sector of the SM  after the
 quark and
gluon degrees of freedom have been integrated out.  The form of
this effective field theory and all its possible terms
are determined by $SU_L(3) \times SU_R(3)$
chiral invariance and Lorentz invariance.
Terms which explicitly break chiral invariance are introduced in terms
of the quark mass matrix 
${\cal M} = \mbox{diag} [ m_u, m_d, m_s ] =\cal{M}^\dag$. 
The lowest order strong chiral lagrangian is 
\beq
{\cal L}_{\rm strong}^{(2)} = 
\frac{f^2}{4} \Tr \left( D_\mu \Sigma D^\mu \Sigma^{\dag} \right)
 +  \frac{f^2}{2}
B_0 \Tr \left( {\cal M} \Sigma^{\dag} +  \Sigma {\cal M}^{\dag} \right) \, .
\label{L2strong}
\eeq
The parameter  $B_0$ is defined by
$\langle \bar{q}_i q_j \rangle = - f^2 B_0 \delta_{ij}$, where
\bea
(m_s + m_d) \langle \bar{q} q \rangle   =  - f_K^2 m_K^2  \; , \qquad
(m_u + m_d) \langle \bar{q} q \rangle   =  - f_\pi^2 m_\pi^2  \; ,
\label{PCACon}
\eea
in the PCAC limit.
The $SU_L(3) \times SU_R(3)$ field 
\beq
\Sigma \equiv \exp \left( \frac{2i}{f} \,\Pi (x)  \right)
\label{sigma}
\eeq
contains the pseudoscalar octet 
  (the would-be Goldstone bosons $\pi$,
$K$, $\eta$)
 $\Pi (x) = \sum_a \lambda^a \pi^a (x) /2 $,
$(a=1,\ldots\,8)$. The coupling
$f$ is, to lowest order,  identified with the  pion decay constant $f_\pi$
(and equal to $f_K$ before chiral loops are introduced); it defines a
characteristic scale
$\Lambda_{\chi} \equiv 2 \pi \sqrt{6/N_c} \, f \, \simeq$ 0.8 GeV,
at which chiral symmetry  breaks down.
 When the
matrix $\Sigma$ is expanded in powers of $f^{-1}$, the lowest
 order term is the 
free  Klein-Gordon lagrangian for the pseudoscalar particles.

To  next-to-leading order there are ten terms and with
 coefficients
$L_i$ to be determined  experimentally or by means of some
model. Some of them play an important role in the physics 
of \ee. As an example, we display the $L_5$ and $L_8$ terms in
${\cal L}_{\rm strong}^{(4)}$ governing much of the
penguin physics:
\begin{eqnarray}
{\cal L}_{\rm strong}^{(4)} \; = \; ..... \; + \,
L_5 \, B_0 \, \Tr \left[ D_\mu \Sigma D^\mu \Sigma^{\dag} 
 \left( {\cal M} \Sigma^{\dag} +  
\Sigma {\cal M}^{\dag} \right) \right] \, + .....  \nnu \\
+ \, L_8 \, B_0 \, \Tr \left[ {\cal M}^{\dag}
 \Sigma  {\cal M}^{\dag} \Sigma +
 {\cal M} \Sigma^{\dag}  {\cal M} \Sigma^{\dag} \right] \, + \, ......
\label{L4strong}
\end{eqnarray}

\subsection{Bosonization of Currents and Densities}
Using factorization (VSA) of currents and densities in the four quark
operators $Q_2$ and  $Q_6$ (as typical examples), one obtains
\bea
\langle \pi^+ \pi^-|Q_2| K^0 \rangle  = 
 - \, \langle \pi^-|\overline{u}\gamma^\mu \gamma_5 d|0 \rangle
\langle \pi^+|\overline{s} \gamma_\mu \, u |K^0 \rangle \; , 
\hspace{2cm}
\label{MQ2} \\
\langle \pi^+ \pi^-|Q_6| K^0 \rangle \; = \; 
 2 \, \langle \pi^-|\overline{u}\gamma_5 d|0 \rangle
\langle \pi^+|\overline{s} u |K^0 \rangle \, - \, 
2  \langle \pi^+ \pi^-|\overline{d} d|0 \rangle  \langle 0|\overline{s} 
\gamma_5 d |K^0 \rangle
\nnu \\
 + 2 \, 
\left[ \langle 0|\overline{s} s|0 \rangle \, - \, 
\langle 0|\overline{d}d|0 \rangle
\right] \,   \langle \pi^+ \pi^-|\overline{s}\gamma_5 d |K^0 \rangle \; ,
\hspace{2cm}
\label{MQ6}
\eea
where the matrix elements  of the densities 
$\overline{s} \gamma_5 u$ and
$\overline{s} d$ might be obtained
from PCAC from matrix elements of (vector and axial) currents.
Let us stress that there is no theoretical underpinning for the VSA; it
is just a convenient reference point which is equivalent
 to retaining terms of
$O(1/N_c)$ in the $1/N_c$-expansion, $N_c$ being the number of colours.

 From (\ref{L2strong}) and  (\ref{L4strong}) we find the bosonized currents
and densities:
\bea
\overline{q}_L^j \gamma^\mu q_L^i \; \rightarrow
 -i \frac{f^2}{2} \left[  \Sigma^{\dag}D^\mu \Sigma \right]^{ij} \; , 
\hspace{2cm}
\label{BosC}  \\
\overline{q}_L^j  q_R^i \; \rightarrow
 - 2 B_0 \left[\frac{f^2}{4} \Sigma \, + \, 
L_5 \, \Sigma D^\mu \Sigma^{\dag}  D^\mu \Sigma \; + \;
 2 L_8 \,  \Sigma  {\cal M}^{\dag} \Sigma 
 \right]^{ij} 
\label{BosD}
\eea
where $i,j$ are flavour indices ($u,d,s$).

Using Fierz transformations and properties of 
the $SU(3)_c$ colour matrices $T^a$, 
the operator $Q_1$ (and similar for other operators) can be rewritten
\bea
Q_{1}  \; = \; \frac{1}{N_c} \,  Q_2 + 
 8 \;  ( \overline{s}_L \, \gamma^\mu \, T^a \, u_L \, ) \; 
( \overline{u}_L \, T^a \, \gamma_\mu \, d_L \,) \; .
\label{FiCo}
\eea
The last term in (\ref{FiCo})  corresponds to colour
exchange between two currents and 
is genuinely non-factorizable.

\subsection{The Weak Chiral Lagrangian}

 The $\Delta S=1$ chiral lagrangian at $O(p^2)$ might be written as
 nine terms, of which we explicitely display four:
\bea
{\cal L}^{(2)}_{\Delta S = 1}  = && G_{LR}^{(0)}(Q_{7,8})
\Tr \left( \lambda^3_2 \Sigma^{\dag} \lambda^1_1 \Sigma
\right) \; + \; 
 G_{\underline{8}} (Q_{3-10}) \Tr \left( \lambda^3_2 D_\mu \Sigma^{\dag}
D^\mu \Sigma\right)   \nnu \\
&+& G_{LR}^{(m)} (Q_{7,8})\ \left[ \Tr \left( \lambda^3_2 
\Sigma^\dag \lambda _1^1 \Sigma {\cal M}^\dag \Sigma \right)
+ \Tr \left(\lambda_1^1 \Sigma  \lambda^3_2  \Sigma^\dag 
{\cal M} \Sigma^\dag \right)\right] \nnu \\ 
&+&  G_{LL}^a (Q_{1,2,9,10}) \, 
\Tr \left(  \lambda^3_1 \Sigma^{\dag} D_\mu \Sigma \right)
\Tr \left( \lambda^1_2 \Sigma^{\dag} D^\mu  \Sigma \right) \; + \, ...
\label{Lchi} \, ,
\eea
where $\lambda^i_j$ are combinations of Gell-Mann
$SU(3)$ flavour matrices defined by
 $(\lambda^i_j)_{lk} = \delta_{il}\delta_{jk}$
and $\Sigma$ is defined in \eq{sigma}.
In (\ref{Lchi}) we  have indicated the quark operators
from which the $G_j$'s the get their contrubution.

 In the leading order in
$1/N_c$  the two most important contributions to \ee might be found 
 by using
 (\ref{Lquark}, \ref{BosC}, \ref{BosD}).
\beq
 G_{\underline{8}} (Q_{6}) = - 24 
\frac{\langle \bar q q \rangle ^2 L_5}{f^2} \, C_6 \; , \qquad
G^{(0)}_{LR}(Q_{8}) = - 3 \langle \bar q q \rangle ^2 \, C_8 \; .
\eeq
\section{Theoretical Predictions}

The generic amplitude for $K^0$ 
to decay into two pions is
\bea
\langle ( \pi \pi )_{(I)} | {\cal L}_W | K^0 \rangle & =& - i A_I 
\, \exp \: (i \, \delta_I) \; , 
\label{def2}
\eea
where the phases $\delta_I$ come from the pion final-state 
interactions (FSI).
The decay of $\bar K^0$ is given by the same expression, except that
$A_I$ is replaced by its complex conjugate $A_I^*$.
 The phase is the same according to 
Watsons theorem. 
The smallness of 
 $\omega \equiv (\Re A_2/\Re A_0) \simeq 1/22.2$ is the 
 $\Delta I = 1/2$  rule of $K\to \pi\pi$ 
decays. Any approach used to predict \ee should also reproduce this
important selection rule. \ee can be written
\beq
 \frac{\varepsilon '}{\varepsilon} =  
\frac{G_F}{2\mod{\epsilon}\Re{A_0}} \:
\mbox{Im}\, \lambda_t \: \:
 \left[\omega \, \Pi_0 -  \Pi_2 \right] \; ,
\label{epsprime2}
 \eeq
where, referring to the $\Delta S=1$ quark lagrangian of \eq{Lquark},
\bea
\Pi_I =  
\frac{1}{\cos\delta_I} \sum_i y_i \, 
\Re  \langle ( \pi \pi )_{(I=0,2)} | Q_i | K^0 \rangle
\label{Pi_I}
\eea
for isospin $I=0,2$ for the pions. For $\Pi_0$, there is also an isospin
breaking factor $(1-\Omega_{IB})$ attached to the right hand side 
of (\ref{Pi_I}). The value 0.25 was used for 
$\Omega_{IB}$. Recent analysis
give shifted values (see refs. 33-36 in \cite{BEFup}).

Notice that  $\omega \, \Pi_0$ (dominated by $Q_6$) has the same sign
and is, within the VSA, almost of the same order of magnitude
 as $\Pi_2$ (dominated by $Q_8$). 
Still,  $Q_6$ gives the biggest contribution to \ee. For instance,
the $Q_6$ contribution is enhanced by chiral loops.

Notice also the explicit presence of the final-state-interaction phases
in eq. (\ref{Pi_I}). 
Their presence is a consequence of writing the absolute values 
of the amplitudes in term of their dispersive parts. 
A fit from the experimental data gives
$\delta_0 \simeq 37^0$ and   $\delta_2 \simeq -7^0$,
in agreement with $\chi PT$.
FSI therefore enhances the $I=0$ over the $I=2$ amplitude
by about 20\%.
(This effect is not explicitly included in some existing
 estimates.)

\subsection{The Chiral Quark Model ($\chi QM$)}

This model has the following  term  added to the QCD lagrangian \cite{ChQM}:
\beq
 {\cal{L}}_{\chi \mbox{\scriptsize QM}} = - M \left( \overline{q}_R \; \Sigma
q_L +
\overline{q}_L \; \Sigma^{\dagger} q_R \right) \, ,
\label{M-lag}
\eeq
which introduces meson-quark couplings.
 The quantity $M$ is
intepreted as the constituent quark mass, appearing because of chiral
 symmetry breaking.

Within the $\chi QM$, the matrix elements in (\ref{MQ2}, \ref{MQ6})
 are  evaluated to $O (p^4)$ within the model. The current matrix elements
of lowest order are  well known,
 and the $\chi$QM version of these 
 (given by divergent quark loops),
 are in agreement with these
by construction.
Futhermore, the $L_i$'s of (\ref{L4strong}) can be calculated, for instance
\beq
L_5 = - \frac{f^4}{8 {\langle \bar{q} q \rangle }} \frac{1}{M} \left( 1 - 6
\frac{M^2}{\Lambda_\chi^2}\right) \, .
\label{L5}
\eeq
The model has a  ``rotated  picture'', where
 the term ${\cal{L}}_{\chi QM}$ in (\ref{M-lag}) is transformed into a pure
mass term $- M \overline{\chi} \chi$
for rotated "constituent quark fields'' 
 $\chi_L =  \xi  q_L$ and 
$\chi_R =  \xi^\dagger q_R$,  
where $\xi \, \cdot \, \xi  = \Sigma $.
The meson-quark couplings in this rotated  picture  arise from
the kinetic part of the constituent quark lagrangian.
These interactions can be described in terms of vector and axial vector
fields coupled to the constituent quark fields
The axial field, being invariant under local chiral transformations, is
given by
\beq
2 i {\cal{A}}_{\mu} \, = \,  \xi^\dagger \, \partial_{\mu} \xi  
 \; - \; \xi \,  \partial_{\mu} \, \xi^\dagger  \, =  \, 
- \xi^{\dagger} (D^{\mu} \Sigma) \xi^{\dagger} \; = \;
 \xi (D_{\mu} \Sigma^{\dagger}) \xi  \, . 
\label{Afield}
\eeq
Using (\ref{Afield}), the  strong chiral lagrangian  $O (p^2)$
 can be understood
as two axial fields ${\cal{A}}_{\mu}$ coupled to a quark loop, giving
${\cal L}^{(2)}_{s} \, \sim \,  \Tr \Bigl[ {\cal{A}}_{\mu} \, {\cal{A}}^{\mu}
 \Bigr] $.
 The QCD current mass lagrangian 
 can be transformed to the form
\beq
{\cal{L}}_{cm} \, = \, - \, \overline{\chi}
 \left( \widetilde M_q^V \, + \, \gamma_5 \, \widetilde M_q^A \right)
 \chi \; , \quad
 \widetilde M_q^{V,A}   \; \equiv  \;\frac{1}{2} \left(\xi^{\dagger} \,
 {\cal{M}}_{q}
  \xi^{\dagger} \,    \pm   \,  \xi \,
{\cal{M}}_{q}^{\dagger} \,  \xi \right) \,  \, .
\label{curmass}
\eeq
Similarly, a lefthanded current can be written ($\lambda^X$ is a SU(3)
flavour matrix)
\beq
\bar q_L \gamma^\mu \lambda^X q_L \, = \,  \,
\bar \chi_L  \gamma ^\mu \Lambda^X  \, \chi_L  \; ;
\qquad
\Lambda^X \, \equiv \, \xi \lambda^X \,  \xi^{\dagger} 
\label{leftcur}
\eeq
By coupling the fields ${\cal{A}}_{\mu},\, \widetilde M_q^{V,A}, \,
\Lambda^X$ to quark loops, the chiral lagrangians
 in sect. 2.2-4, and higher order terms can be calculated.

Within the $\chi QM$ one can  include soft gluon contributions
corresponding to the nonfactoralizable term in (\ref{FiCo}).
One calculates soft gluon emision from the two coloured currents,
and identify the product of two gluon field tensors  with 
the gluon condensate $<G^2>$. As an example \cite{PdeR,BEF96}:
\bea
G_{LL}^a(Q_1) = \frac{1}{N_c} \; f_\pi^4 \, \left( 1 - \delta_G \right)
\; \; , \qquad
\delta_G \equiv  N_c \, \frac{<\alpha_s GG/\pi>}{32 \pi^2 f^4} \; .
\label{GLLa}
\eea
The hadronic matrix elements of $Q_i$ in (\ref{Pi_I}) to $O (p^4)$ 
has three terms: 

i) A three level term obtained from the
$O (p^2)$ chiral lagrangian  (\ref{Lchi}). 

ii) The chiral loops
contributions from (\ref{Lchi}), being $O (p^4)$ .

iii)  $O (p^4)$ terms obtained by calculating the matrix elements
in (\ref{MQ2},\ref{MQ6})  up to $O (p^4)$ (as we did in 
\cite{BEF96,BEFdi,BEFee,BEFup}),
 or equivalently, one may  calculate
the $G_i$'s of the $O (p^4)$ weak lagrangian.

\begin{figure}
\begin{center}
\epsfig{figure=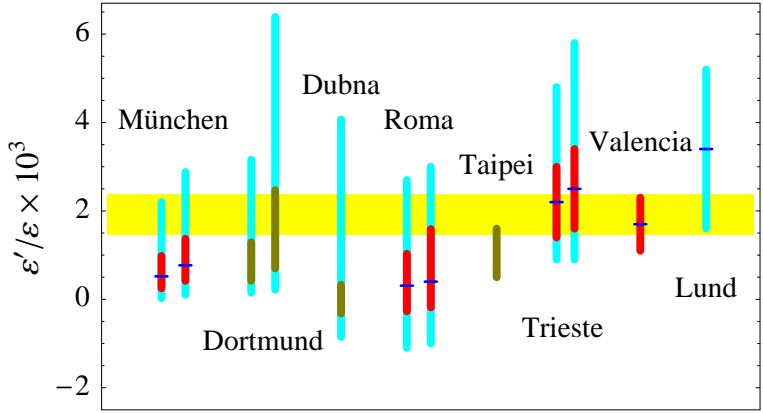,width=10cm}
\caption{Summary of theoretical estimates for \ee. The dark (light)
 range corresponds to Gaussian (flat) scanning of the uncertainties.
The gray horisontal band is the average experimental result.}
\label{theorest}
\end{center}
\end{figure}

 The hadronic matrix elements have a scale dependence
from the chiral loops which is mathched to the one in the 
 Wilson coefficients
at the scale $\mu = \Lambda_\chi \simeq 0.8$ GeV.
We find \cite{BEFdi}  that the $\Delta I = 1/2$ rule is well reproduced
for  the reasonable values (see also \cite{PdeR}. Notice 
that $\delta_G \simeq 2$.) 
\beq
M \simeq  200  \mbox{MeV} \, ,
\quad
 \langle \alpha_s GG/ \pi \rangle \simeq  
\left( 330 \mbox{MeV} \right) ^4 \, , \quad
\langle \bar q q \rangle \simeq
 \left( -240  \mbox{MeV} \right) ^3 \; .
\label{range-MP} 
\eeq
These   are used as input to estimate \ee, and we
got \cite{BEFee} a result close to the one which later
 became the world average $\Re \: \varepsilon '/\varepsilon  = 
(19  \pm 2.4)  \times   10^{-4} \;$, based on
 results from 1992 at
 CERN and FNAL, and  the last  years run at FNAL (KTeV) \cite{KTeV} and 
CERN (NA48) \cite{NA48}.
Our (``Trieste group'')
estimate is shown together with other theoretical estimates in
Fig. \ref{theorest}.
For  precice statements about numerical values, I refer to 
the cited papers.

The strength of our approach \cite{BEF96,BEFdi,BEFee} 
is that all contributions are calculated
systematically up to $O(p^4)$ in the chiral expansion.
The weakness is its model dependence, which it shares more or less with
other estimates.
A matching of 
SD and LD calculations at $\mu \simeq$ 0.8 GeV has been questioned.
Still, the obtained numerical stability in the matching is good.

\subsection{Other Approaches} 
\begin{itemize}
\item
 \underline{M\"unchen}\cite{Bosch99}: The $\Delta I = 1/2$ rule is fitted
(at $\mu=m_c$). Unfortunately, this does
 not give (enough) information about the
 matrix elements of $Q_{6,8}$. These are varied  around their leading 
$1/N_c$ values.

\item
 \underline{Roma}\cite{Rom}: This is  the only first principle
estimate. Some matrix elements have been calculated, but unfortunately
not that of $Q_6$, which is 
varied by 100\% around its VSA value. 
Also the ``penguin contractions''
of $Q_{1,2}$ needed to explain the $\Delta I = 1/2$  rule
are difficult to  obtain.

\item
 \underline{Dortmund} \cite{Dort}: The quark operators are matched
(between 600 and 900 MeV)
 directly to their corresponding chiral loops with a quadratic cut-off,
carefully identified with the renormalization scale.

\item
 \underline{Dubna}\cite{Dub}: Uses extended Nambu Jona-Lasinio
(ENJL) models
and chiral loops regularized by the heat kernel. It is found that
chiral loops to $O(p^6)$ have sizeable absorptive parts bur rather
 small real parts.

\item
 \underline{Taipei}\cite{Tai}: Uses ``generalized factorization'' where
scheme dependence is absorbed in effective Wilson coefficients.
Non-factorizable contribtions in (\ref{FiCo},\ref{GLLa}) are represented
by a phenomenological parameter.

\item
 \underline{Valencia}\cite{Val}: Special attention is put on FSI effects.
A dispersion relation is used, and the analysis confirms the importance 
of chiral loops.

\item
 \underline{Lund}\cite{Lun}: A EJNL framework including  
axial and  vector resonances. Care is taken to match LD and SD calculations.
A sizeable enhancement of
the $Q_6$ contribution is found, which predicts  \ee on the high side. 

\end{itemize}
Also, an estimate using a  linear sigma model \cite{KNS} shows
enhancement of the $Q_6$ matrix element, but its hard to explain
the $\Delta I = 1/2$ rule and \ee at the same time within this approach.

\section{Conclusions}

 All groups have to a certain extent  used 
assumptions and models. Still, most estimates are roughly in the right
ballpark, as seen from Fig. \ref{theorest}. Given the hadronic uncertainties,
it will be hard to disentangle  new physics from the SM predictions.

\begin{center}
   * \qquad * \qquad *
\end{center}

I thank the organizers of this meeting, and my collaborators
 S. Bertolini  and M. Fabbrichesi.


\end{document}